# Modelling of microstructures during in-situ alloying in additive manufacturing for efficient material qualification processes

*Modellierung von Mikrostrukturen während des in-situ Legierens in der additiven Fertigung für effiziente Materialqualifizierungsprozesse*

Patrick Zimbrod, Johannes Schilp, Universität Augsburg, Augsburg (Germany),
patrick.zimbrod@informatik.uni-augsburg.de, johannes.schilp@informatik.uni-augsburg.de

**Abstract:** In this work, a numerical simulation framework is presented based on the Phase Field Method that is able to capture the evolution of heterogeneous metallic microstructures during solidification. The involved physics can prove especially useful when studying not only systems undergoing thermal gradients, such as in homogeneous systems, but also in conditions that exhibit stark spatial gradients, i.e. when these inhomogeneities are present even on a mesoscopic scale. To illustrate the capabilities of the model, in-situ alloying of a High Entropy Alloy during Laser Powder Bed Fusion is investigated as an exemplary use case. The resulting digital twin is expected to shorten development times of new materials as well as cut down on experimental resource needs considerably, therefore contributing to efficient material qualification processes.

## 1    Introduction

The phenomenological modelling of phase evolution dates back to over 70 years and has been used widely in computational material science ever since (Cahn and Hilliard 1958). The method has especially proven its worth in the analysis of alloy solidification which has been well established for binary alloys (J. A. Warren and Boettinger 1995). With increasing computational resources, this approach has also been adapted to more demanding problems such as the solidification of polycrystalline materials (James A. Warren et al. 2003). Since then, phase field modelling has undergone a considerable rise in popularity within material science such that it is almost impossible to extensively cover all applications of this method (Steinbach 2009).

Because of the valuable insight that modelling highly dynamic systems in thermodynamic non-equilibrium provides, the analysis of microstructures within additive manufacturing has also been subject to extensive efforts by various authors (DebRoy et al. 2018). Within powder bed fusion additive manufacturing, especially the micro-scale coalescence behaviour of the present phases is of interest and has therefore been investigated (Shinagawa 2014). Beyond that, because of the stark thermal gradients typically present in those processes, the morphology of the evolving microstructure is often of particular interest, limiting the mechanical performance of the as-built alloys (Ji, Chen, and Chen 2017).

Within the normal process of powder bed fusion though, pre-alloyed powders are fed into the process and molten by a laser or electron source, yielding a comparably homogeneous distribution of the alloying elements. In contrast, in-situ alloying relies on mixing the pure elemental powder in the melting process itself (Katz-Demyanetz, Koptyug, and Popov 2020). This method naturally causes local gradients with respect to the alloy species that have to be included in the modelling approach. Therefore, additional sub-models are required in the form of partial differential equations (PDE) in order to be able to solve a closed system. Additional information on this topic is given in section 3.

Hence, we develop a computational framework that covers the physical effects of these gradients in addition to the nonetheless present polycrystalline solidification. It is anticipated that this will provide a valuable framework for preliminary case studies in material qualification. This can greatly assist the aim of digitally supplemented discovery of new materials, which has already been discussed in the literature (Megahed et al. 2019). Through in-situ alloying, costly powder alloying steps can potentially be further reduced and elemental powders more universally been used (see Fig. 2). Consequently, the experimental effort could be shifted to simulative studies to some extent, hence reducing the use of valuable resources in production. This in turn has also been previously identified as a core need for modern material qualification in additive manufacturing (Seifi et al. 2016).

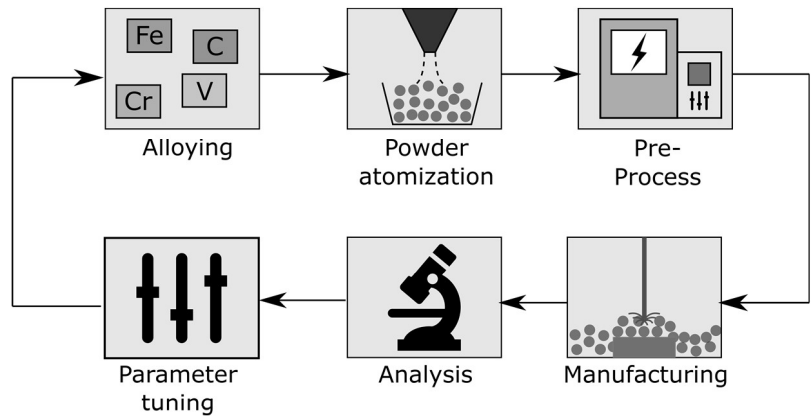

*Figure 1:* A simplified illustration of the typically strongly empirical and iterative material qualification process within powder bed additive manufacturing. Several pieces of specialized equipment along with substantial process knowledge is required for reliable operation. In-situ alloying typically renders the first step of the process obsolete as alloying occurs during the manufacturing step. The total amount of iterations of this process for a new material is expected to be reduced by adding the proposed digital twin approach.

In summary, in-situ alloying can be considered a logical next step regarding flexible materials discovery. By providing a computational framework to simulate microstructure evolution in these conditions, we aim to support and extend the digital material qualification by an enhanced digital twin that enables a fast discovery of novel candidates for high-performance alloys.

## 2　　Previous Works

In-situ alloying as a means of accelerated parts qualification and increased flexibility in production has spiked interest within additive manufacturing in the recent years (Katz-Demyanetz, Koptyug, and Popov 2020). As described previously, this approach introduces additional variability in the process that has to be addressed.

Ma et al. have conducted preliminary studies on how local inhomogeneities of this type affect microstructures in titanium using tungsten arc welding (Ma et al. 2015). Simonelli et al. investigated resulting microstructures from different kinds of powder blending. They found that a suitable arrangement of particle size distribution (PSD) between the alloying elements can impact the mechanical performance positively (Simonelli et al. 2018).

Especially the development of completely new alloys has been making use of in-situ alloying as a cost-effective technique for Design of Experiments type studies. A prominent example is the group of High Entropy Alloys (HEA) that have picked up considerable attention within additive manufacturing (Torralba and Campos 2020). Among those, several studies have been conducted using in-situ alloying, solidifying its worth in material qualification (P. Chen et al. 2020; Kim, Yang, and Lee 2020; Pegues et al. 2020; Sun et al. 2020; Gao et al. 2020).

The modelling and simulation in turn has mainly been focusing on predicting the microstructures of homogeneous powder blends. Using the phase field method in powder bed fusion, several authors provided a foundation for simulating grain evolution, mostly being coupled to mesoscale simulations to obtain the necessary information about the temperature fields (Liu and Shin 2020; Staroselsky, Acharya, and Cassenti 2020; Zielinski et al. 2020; D. Chen, Wang, and Zhang 2021; Wang et al. 2021). Therefore, it is considered a valuable addition to the current research to introduce spatial gradients into the analyses with the aim of modelling the complex physical conditions present in in-situ alloying. In particular, the investigation of inhomogeneous fields dictates the use of methods that track additional quantities in the phase field model. One way to incorporate this complexity is by using species-specific order parameters that are solved simultaneously. This so called multiphase field method has been applied successfully to solidification problems (Schmitz and Prahl 2016). However, information on orientation can hardly be tracked as each order parameter would be coupled to an individual orientation variable. Furthermore, diffusion between the phases cannot be considered which has proven to be important in form of interstitial crystal formation (Moghaddam et al. 2020). Hence, a more refined approach to this problem is needed.

## 3　　Phase field modelling and the CALPHAD method

The starting point of a phase field model generally is the construction of a functional that describes the physical system in terms of an extensive thermodynamic variable.

This expression must commonly follow a minimum principle in order to represent an equilibrium state and can consequently be used in order to obtain equations for the spatial and temporal evolution of the system.

One can construct these equations by taking the variational derivatives of the functional in terms of the dependent variables. A characteristic feature of the phase field method is the use of a so-called phase variable or order parameter, typically denoted with $\phi$. This has proven to be a useful approximation for systems undergoing solid phase transformations, as solids tend to be characterized by internal ordering of some degree (Provatas and Elder 2010). Here, we will consider the Helmholtz free energy functional $\mathcal{F}$ which has been a popular choice besides the entropy functional (Provatas and Elder 2010). This expression for this particular application depends on the phase variable $\phi$, crystallographic orientation $\theta$ as well as the species concentrations $C_j$ of the alloying elements.

The formulation of the Helmholtz functional that is based on the works of Warren et al is given by (James A. Warren et al. 2003):

$$\mathcal{F} = \iiint_\Omega \left\{ f(\phi, C_j, T) + \frac{\alpha^2}{2} \Gamma(|\nabla \phi|, \theta - \psi) + sg(\phi)|\nabla \theta| \right. \\ \left. + \frac{\epsilon^2}{2} h(\phi)|\nabla \theta|^2 \right\} dV \qquad (1)$$

We now substitute the free energy density $f$ with a more specific thermodynamic formulation that incorporates the variation of alloy species into the model:

$$f(\phi, C_j, T) = \sum_{j=1}^{N} C_j \left[ \mu_j(\phi, T) + RT \ln\left(\frac{C_j}{\rho}\right) \right] \qquad (2)$$

In this way, we can capture the influence of changing species concentrations as well temperature dependence through the chemical potentials in the domain. Though, we now need the respective chemical potentials $\mu_j$ as a function of phase and temperature as inputs. Furthermore, we have to track $N$ more quantities, namely the respective species concentrations. From this point, we can compute the variational derivatives $\delta \mathcal{F}$ in terms of the dependent variables and obtain the governing equations for the problem. For sake of brevity, the details of this formulation will not be discussed here but instead further detailed in a future publication. The equations for an uncoupled system containing the variables given above are given in the literature (James A. Warren et al. 2003; Guyer et al. 2004). Computing the variational derivatives $\delta \mathcal{F}/\delta \phi$ and $\delta \mathcal{F}/\delta C_j$ using the free energy formulation in Eq. 2 allows to track the evolution of interstitial and substitutional species in the solidified alloy system, which is highly relevant for in-situ alloying, as discussed in previous sections.

In order to obtain the chemical potentials $\mu_j$ that serve as inputs in Eq. 2, one can employ the computational calculation of phase diagrams approach (CALPHAD). Although being heavily reliant on thermodynamic databases, this style of modelling is capable of calculating equilibrium states and quantities at given process conditions. The basic idea can be illustrated by constructing a mixed Gibbs free energy formulation by using the pure components $G_{A/B}$, an interaction term $L_{AB}$ and the thermodynamic conditions, e.g. temperature $T$ and compositions $x_{A/B}$ (Czichos, Saito, and Smith 2006):

$$G_m = x_A\,G_A + x_B G_B + x_A x_B L_{AB} + RT(x_A \ln x_A + x_B \ln(x_B)) \qquad (3)$$

The calculation of other thermodynamic quantities follows a similar, straightforward approach and is well-documented in the literature (Czichos, Saito, and Smith 2006).

## 4      Methodology

The phase field model in the present formulation contains various physical and numerical parameters that are listed in Table 1. Further details on the reasoning behind these constants as well as the numerical choice are given in the original publication by Warren et al. (James A. Warren et al. 2003). These among others control the physical effects that are present during solidification, e.g. grain rotation, anisotropy leading to dendritic growth, grain coalescence and coarsening or neck growth kinetics.

*Table 1: Choice of mostly non-alloy-specific modelling parameters for the present phase field simulation (James A. Warren et al. 2003)*

| Quantity | Description | Value | Unit |
|---|---|---|---|
| $L$ | Volumetric latent heat | 2 e9 | $[J/m^3]$ |
| $a$ | Double well potential height | 31,6 e3 | $[(J/m^3)\wedge 0,5]$ |
| $\epsilon$ | $\phi$-$\theta$ interaction energy coefficient | 44,7 e3 | $[(J/m^3)\wedge 0,5]$ |
| $\alpha$ | $\phi$ gradient energy coefficient | 83,9 e3 | $[(J/m^3)]$ |
| $s$ | $\theta$ gradient energy coefficient | 4,69 e9 | $[J/m^3]$ |
| $c$ | Additional heat source coefficient to stabilize domain temperature | 0,1 | $[1/s]$ |
| $N$ | Order of crystal anisotropy | 4 | $[-]$ |
| $\tau_\phi$ | Kinetic phase scaling factor | 3 e-4 | $[-]$ |
| $\tau_\theta$ | Kinetic orientation scaling factor | 3 e5 | $[-]$ |
| $D_T$ | Thermal diffusivity coefficient | 2,25 | $[m^2/s]$ |
| $\kappa_1$ | Free energy phase scaling coefficient | 0,9 | $[-]$ |
| $\kappa_2$ | Free energy temperature scaling coefficient | 20 | $[-]$ |

The proposed digital qualification process follows the procedure depicted in Fig. 2. After information about composition, temperature and thermodynamics are gathered from CALPHAD and Computational Fluid Dynamics calculations, the phase field model tracks the evolution of the microstructure being investigated in terms of the order parameter, crystallographic orientation and species concentration.

As a result, we obtain a two-dimensional phase and orientation field that can be colour-mapped to graphically encode the grain orientation and boundaries. The morphology of the grains can then be compared to Inverse Pole Figures (IPF) from experimental data.

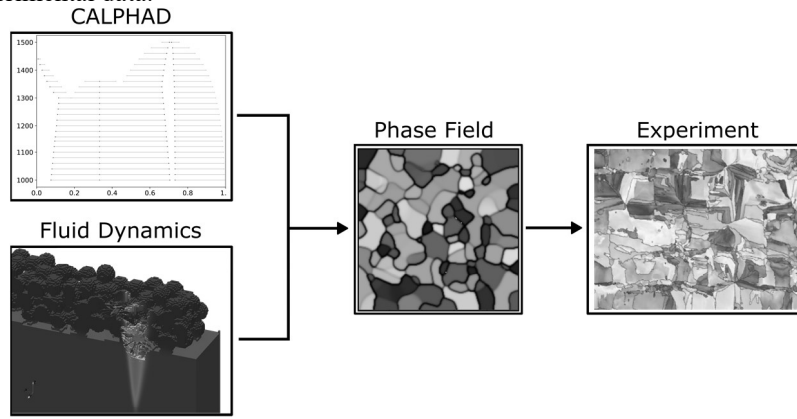

*Figure 2: Illustration of the proposed computational materials engineering workflow: Thermodynamic calculations alongside with thermo-fluid-dynamics modelling provide inputs for microstructure evolution models that can be directly evaluated.*

Using the computational methods presented in the theory, we create the relevant chemical potentials for the investigated quantities using the CALPHAD approach. Within this study, we used the *pyCALPHAD* library alongside with the openly available thermodynamic database for nickel-based alloys for our calculations (Otis and Liu 2017; Povoden-Karadeniz 2016).

In order to solve the phase field problem numerically, we employ the finite volume library *fipy* alongside with the *PETSc* numerics suite and its respective Python bindings (Guyer, Wheeler, and Warren 2009; Balay et al. 2019; Dalcin et al. 2011). We use simple upwinding as the flux function and solve the problem using a Jacobi-preconditioned linear solver based on the generalized minimal residual method (GMRES) for its numerical robustness and reasonable weak scaling among many processors.

## 5    Results and Discussion

As outlined in section 4 and illustrated in Fig. 2, we need thermodynamic data in order to appropriately incorporate the influence of species into the simulation. Otherwise, the first contribution to the integral in Eq. 1 would not be representative. The calculated chemical potentials that serve as input for the phase field model (Eq. 1) are given in Figure 3. These results represent the numerical values of $\mu_j(\phi, T)$ that were introduced in Eq. 2 and can be used directly to compute the evolution of the species concentrations.

Regarding the validity of CALPHAD calculations, it has to be noted that this method is of purely data-driven nature. The framework in itself only provides rules to interpolate given values for pure elements in a thermodynamically consistent manner. It is therefore crucially important to rely on valid data in order to create consistent results (Otis and Liu 2017). For this reason, we chose the open database offered by Povoden-Karadeniz as it contains large amounts of consolidated data from well over 200 references (Povoden-Karadeniz 2016). Despite these obvious limitations,

CALPHAD has proven to be a highly useful and relied on tool to create thermodynamic quantities for arbitrary alloys that are valid within the defined boundaries of the database considered.

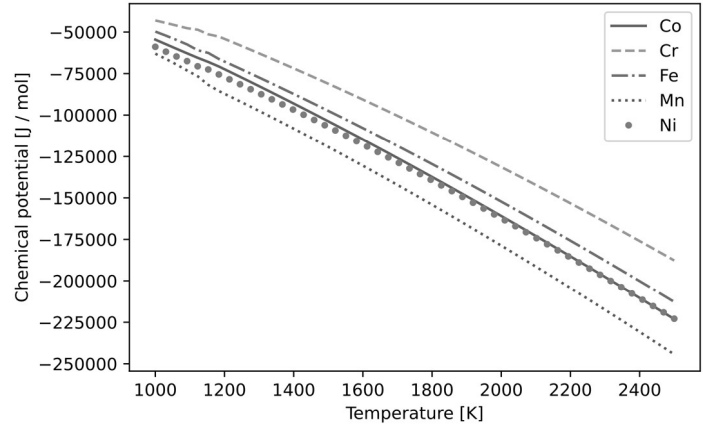

**Figure 3:** *Chemical potentials $\mu_j$ of the investigated alloying elements between 1000 K and 2500 K. Calculations were performed using the CALPHAD method*

Going on, we take the constant values given in Table 1 and assemble the remainders of the model. In this case, we simplify the input regarding boundary conditions to a constitutional undercooling of 100 K that is often found in the typically harsh process conditions during powder bed fusion (DebRoy et al. 2018). Additionally, we impose Neumann boundary conditions of the form

$$\partial u(x_i) = 0, \qquad \forall x_i \in \partial \Omega \qquad (4)$$

on all walls $\partial \Omega$ to create an adiabatic system. Within the simulated domain of *100 x 100 µm*, we seed 100 grains at a random location and orientation in order to approximate the real grain growth behavior. For this problem, we use a grid size of *0.25 µm*, resulting in a total 160.000 cells.

As the process of nucleation itself is of stochastic nature both with respect to the sites and to the crystallographic orientation, the quantities output by the phase field simulation will inherently not match a given manufactured microstructure precisely. Instead, we will be looking at the grain morphologies, i.e. the shape and size distribution for a given domain. These characteristics are of vital importance in order to estimate the mechanical properties of an alloy system.

A resulting microstructure at varying timesteps is given in Fig. 4. The snapshots clearly show the originally anisotropic, dendritic growth of the initialized grains that over time coalesce and impinge with other grains. The morphology shown is not uniform owing to the present anisotropy and yields grains of noticeably different size. The dendritic arms that stretch out of the initially circular grains show the imposed four-fold anisotropy. Some of the solidified grains end up with an elongated morphology that would deteriorate fracture elongation considerably in a printed part. These phenomena can also be observed in the printed specimens manufactured by Chen et al.

In Fig. 4 d), we observe several different grain sizes and aspect ratios, i.e. the ratio between length and width. Comparing these irregularly shaped grains with the region of interest in Fig. 4 e), we see that the structures share indeed similar characteristics regarding size and morphology. The manufactured structure by Chen et al. contains highly irregular grains that (I) have a very high aspect ratio with some grains having about 3x more length than width and (II) exhibit an almost rectangular shape. Both effects can be observed in Fig. 4 d) as well. Grains of such trapezoidal shape (II) can e.g. be ovserved in the lower right edge of the domain in Fig. 4 d) as well as in the lower left corner for high aspect ratio grains (I). Furthermore, one can also observe typically smaller grains with an aspect ratio much closer to unity, i.e. globular grains in the simulated structure. We therefore conclude that due to these similarities in morphology, the model captures the impingement behaviour of the nuclei in an appropriate manner.

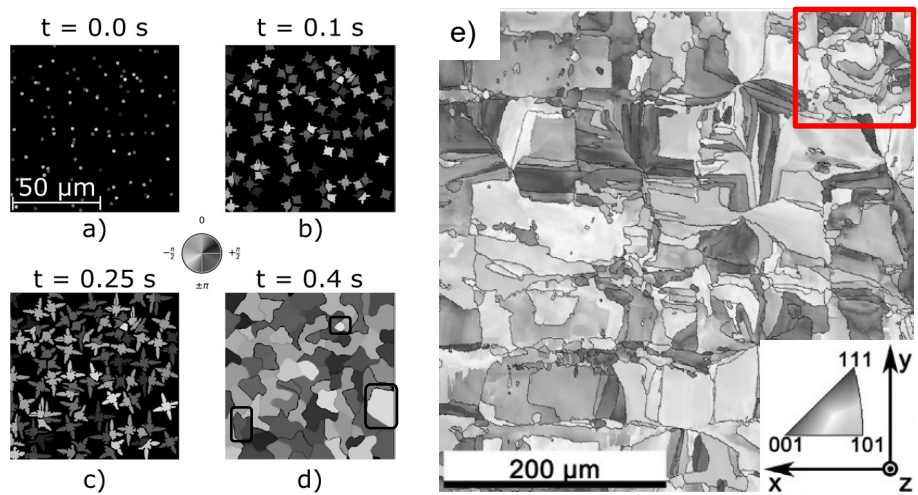

**Figure 4:** *Results of the phase field simulation (a – d) and microstructure of the in-situ alloyed CoCrFeMnNi HEA (e) investigated by Chen et al (P. Chen et al. 2020), where a region of interest matching the simulation domain size is marked in red. One can clearly observe the originally small, circular grains (a) evolving with a dendritic growth pattern (b and c) as well as the subsequent grain impingement that forms the final microstructure (d). Grains of noteworthy morphology are indicated by black squares for illustrative purposes.*

It should be noted that even at the very small scales considered in this study, the computational effort necessary to solve such models is rather large. Negligible resources are needed to obtain the physical quantities from CALPHAD, which for this study took time in the order of few minutes on a four-core mobile processor (Intel® Core i7-8550U). The calculations necessary comprise mainly light I/O operations and well-implemented simple arithmetics, as outlined by Eq. 3. The aforementioned amount of computational expense instead stems from the coupled nature of the governing equations that arise from Eq. 1. On a workstation equipped with an 18-core/36-thread Intel@ Xeon W-2295 paired with 128 GB of DDR4 RAM, the calculations up to a time of 0.4 s took approximately 200 minutes using OpenMPI. Therefore, this type of simulation using the Finite Volume Method is prohibitively expensive for conducting large parameter studies, especially because the parameter space regarding the alloying elements is typically rather large. The same rationale applies for simulating entire layers of a build process since the domain would likely be around two orders of magnitude larger in each direction. One possible next step to remedy this shortcoming could be to conduct these calculations on more parallel architectures such as state-of-the-art GPUs. It has been shown that solving the system of linear equations on GPU architectures can produce faster solutions than on pure CPU architectures (Kuo et al. 2011). However, this speedup diminishes rapidly as soon as the amount of CPU resources grows up to about 48 cores or more. This can be attributed to the rather large MPI communication overhead necessary within the Finite Volume Method. One alternative route would be to transfer the presented phase field method to more efficient numerical schemes regarding parallelization, such as the Discontinuous Galerkin Finite Element Method (DG-FEM). Using this method, much more drastic improvements regarding computational effort can be achieved that can reach up to two orders of magnitude (Kirby and Mavriplis 2020).

## 6  Conclusion

We have proposed a modelling approach alongside a numerical scheme and implementation of modelling in-situ alloying during powder bed fusion additive manufacturing. The model consists of an extended phase field formulation for polycrystalline materials that adds the influence of concentrations and chemical potentials to the free energy formulation. The necessary but typically hard to capture physical quantities are determined using the CALPHAD method. Simulation outputs qualitatively resemble real EBSD graphs that can be captured after processing in-situ alloyed High Entropy Alloys via laser powder bed fusion. This digital twin can therefore be considered a valuable addition to the discovery of new high-performance materials that can be processed via additive manufacturing. Further studies can and should be conducted upon the initial parameter choice, as these impact the quantitative occurrence of grain rotation, coarsening etc. considerably. It has been furthermore outlined that this specific numerical implementation of the phase field method has room for improvement regarding the use in high-throughput parameter studies. As an outlook, two pathways of improving the computational efficiency in the form of GPU parallelization and developing of more advanced numerical schemes have been outlined.